\documentclass[11pt,onecolumn]{ieeeconf}
\usepackage{graphicx}
\usepackage{amssymb,amsmath}
\usepackage{epstopdf}
\IEEEoverridecommandlockouts    

\newcommand{\mbb}[1]{\mathbb #1}
\newcommand{\mbf}[1]{\mathbf #1}
\newcommand{\mcl}[1]{\mathcal #1}

\newcommand{\Rnx}{\mbb{R}^{n_x}}

\newcommand{\R}{\mbb{R}}
\newcommand{\N}{\mbb{N}}

\newcommand{\inte}[1]{\textrm{int}{\mcl{#1}}}

\newcommand{\eBox}{$\hfill\square$}

\newcommand{\mpcLaw}{\kappa}
\newtheorem{theorem}{Theorem}
\newtheorem{definition}{Definition}

\newtheorem{assumption}{Assumption}

\title{\LARGE \bf
Primal or Dual Terminal Constraints in Economic MPC? -- Comparison and Insights
}

\author{Timm Faulwasser$^{1}$ and Mario Zanon$^{2}$
\thanks{$^{1}$TF is with the Institute for Energy Systems, Energy Efficiency and Energy Economics, Department of Electrical Engineering and Information Technology, TU Dortmund  University, 44227 Dortmund, Germany
        {\tt\small timm.faulwasser@ieee.org}}%
\thanks{$^{2}$Mario Zanon is with the IMT School for Advanced Studies Lucca, 55100 Lucca, Italy
        {\tt\small mario.zanon@imtlucca.it}}%
}

\begin{document}
\maketitle

\begin{abstract}

This chapter compares different formulations for Economic nonlinear Model Predictive Control (EMPC)  which are all based on an established dissipativity assumption on the underlying Optimal Control Problem (OCP). This includes schemes with and without stabilizing terminal constraints, respectively,  or with stabilizing terminal costs. We recall that a recently proposed approach based on gradient correcting terminal penalties implies a terminal constraint on the adjoints of the OCP. We analyze the feasibility implications of these dual/adjoint terminal constraints and we compare our findings to approaches with and without primal terminal constraints. Moreover, we suggest a conceptual framework for approximation of the minimal stabilizing horizon length. Finally, we illustrate our findings considering a chemical reactor as an example.

\end{abstract}

\section{The Dissipativity Route to Optimal Control and MPC}

Since the late 2000s, there has been substantial interest in so-called Economic nonlinear Model Predictive Control (EMPC). Early works include \cite{Rawlings09b,Kadam07a}, recent overviews can be found in \cite{Ellis14,Angeli19a,kit:faulwasser18c}. Indeed the underlying idea of EMPC is appealing as it refers to receding-horizon control based on Optimal Control Problems (OCPs) comprising generic stage costs, i.e., more general than the typical tracking objectives. 
In this context, beginning with \cite{Diehl11a,Angeli12a} a dissipativity notion for OCPs received considerable attention.
Indeed numerous key insights have been obtained via the dissipativity route.

 Dissipativity is closely related to optimal operation at steady state \cite{Mueller14a}, under suitable conditions they are equivalent.
In its strict form dissipativity also implies the so-called  turnpike property of OCPs. These properties are a classical notion in optimal control. They refer to similarity properties of OCPs parametric in the initial condition and the horizon length, whereby the optimal solutions stay close to the optimal steady state during the middle part of the horizon and the time spent close to the \textit{turnpike} grows with increasing horizon length. Early observations of this phenomenon can be traced back to John von Neumann and the 1930/40s~\cite{vonNeumann38}. The term as such was coined in \cite{Dorfman58}; it has received widespread attention in economics \cite{Mckenzie76,Carlson91}. Remarkably early in the development of EMPC, the key role of turnpike properties for such schemes had been observed~\cite{Rawlings09b}. However, it took until \cite{Gruene13a}  a first stability proof directly leveraged their potential. Turnpikes are also of interest for generic OCPs in finite and infinite dimensions, see~\cite{Trelat15a,Gruene19a,tudo:faulwasser20g}. 
 
 Strict dissipativity and turnpike properties of OCPs are---under suitable technical assumptions---equivalent \cite{epfl:faulwasser15h,Gruene16a}. The turnpike property also allows to show recursive feasibility of EMPC without terminal constraints~\cite{epfl:faulwasser15g,kit:faulwasser18c}.
Dissipativity can be used to build quadratic tracking costs for MPC~\cite{Zanon14a,Zanon16a,Zanon17b,DeSchutter2020} yielding approximate economic optimality.
It is also related to a positive-definite Gauss-Newton-like Hessian approximation for EMPC~\cite{Zanon20b}.

Strict dissipativity of an OCP allows deriving sufficient stability conditions with terminal constraints \cite{Angeli12a,Diehl11a,Amrit11} and without them \cite{epfl:faulwasser15g,Gruene13a,kit:zanon18a,kit:faulwasser18e_2}.
Dissipativity and turnpike concepts can be extended to time-varying cases \cite{Zanon13d,Zanon17a,Gruene18b} and to OCPs with discrete controls \cite{tudo:faulwasser20f}.
Finally, there exists a close relation between dissipativity notions of OCPs and infinite-horizon optimal control problems \cite{tudo:faulwasser20a}.

This substantial cardinality of results obtained along the dissipativity route to EMPC and OCPs might be surprising at first glance. However, taking into account the history of system-theoretic dissipativity notions---in particular the foundational works of Jan Willems~\cite{Willems72a,Willems72b,Willems71a}---the close relation between both topics is far from astounding. 

Yet, this chapter does not attempt a full-fledged introduction to EMPC, and the interested reader is referred to the recent overviews \cite{Ellis14,Angeli19a,kit:faulwasser18c}. Neither will it give a comprehensive overview of dissipativity notions. 
Instead, here, we focus on a comparison of EMPC schemes with and without terminal constraints. In general, one can distinguish three main classes of dissipativity-based stability approaches:
\begin{itemize}
\item schemes relying on terminal $\{$equality, inequality$\}$ constraints and corresponding terminal penalties, see e.g. \cite{Diehl11a,Angeli12a,Amrit11};
\item schemes using neither terminal constraints nor penalties, e.g. \cite{Gruene13a,epfl:faulwasser15g};
\item schemes avoiding primal terminal constraints and instead using gradient correcting terminal penalties (which imply a dual terminal constraint) \cite{kit:zanon18a,kit:faulwasser18e_2}.
\end{itemize}
Specifically, this chapter compares the three schemes above with respect to different aspects such as the structure of the optimality conditions, primal and dual feasibility properties, and the required length of the stabilizing horizon. 

\vspace*{2mm}
The remainder of this chapter is structured as follows: In Section \ref{sec:Problem} we recall the EMPC schemes to be compared and the corresponding stability results. In Section \ref{sec:Compare} we analyze the schemes with respect to different properties. We also derive formal optimization problems, which upon solving, certify the length of the stabilizing horizon. Section \ref{sec:Examples} will present results of a comparative numerical case study. The chapter closes with conclusions and outlook in Section \ref{sec:Summary}. 

\section{Economic MPC Revisited} \label{sec:Problem}

In this chapter, we consider EMPC schemes based on the following family of OCPs
\begin{subequations}
	\label{eq:original_ocp}
	\begin{align}
	\hspace{-1em}V_N(\hat x_i) \doteq\min_{\boldsymbol{x}, \boldsymbol{u}} \ \ & \sum_{k=0}^{N-1} \ell(x_k,u_k)  + V_\mathrm{f}(x_N) \hspace{-3em} \label{eq:objective} \\
	\text{subject to} \ \  & x_0 = \hat x_i, \label{eq:original_ocp_jc}\\
	& x_{k+1} = f(x_k,u_k), &&\hspace{-0.5em} k\in \mathbb{I}_{[0,N-1]}, \label{eq:original_ocp_dyn}\\
	& g(x_k,u_k) \leq 0, &&\hspace{-0.5em}  k\in \mathbb{I}_{[0,N]}, \label{eq:original_ocp_pc}\\
	& x_N \in \mbb{X}_\mathrm{f}, \label{eq:original_ocp_tc}
	\end{align}
\end{subequations}
where we use the shorthand notation $\mathbb{I}_{[a,b]} := \{a,\ldots,b\}$, $a,b$  integers.
The constraint set for $z\doteq [x~ u]^\top \in\mbb{R}^{n_x+n_u}$ is defined as
\begin{equation} \label{eq:Zset}
\mbb{Z} = \{z \in \mbb{R}^{n_x+n_u}\,|\, g_j(z) \leq 0, \, j \in \mathbb{I}_{[1,n_g]}\},
\end{equation}
where $\mbb{Z}$ is assumed to be compact.  
To avoid cumbersome technicalities, we assume that the problem data of \eqref{eq:original_ocp} is sufficiently smooth, i.e., at least twice differentiable, and that the minimum exists.\footnote{Provided the feasible set is non-empty and compact, existence of a minimum follows from continuity of the objective \eqref{eq:objective}.} 
The superscript $\cdot^\star$ denotes optimal solutions. Occasionally, we denote the optimal solutions
as 
\begin{align*}
\mbf{u}^\star(\hat x_i) &\doteq  \begin{bmatrix}u^\star_0(\hat x_i) ^\top & \dots& u^\star_{N-1}(\hat x_i) ^\top\end{bmatrix}^\top, \\
\mbf{x}^\star(\hat x_i) &\doteq  \begin{bmatrix}x^\star_0(\hat x_i) ^\top & \dots& x^\star_{N}(\hat x_i) ^\top\end{bmatrix}^\top ,
\end{align*}
whenever no confusion about the initial condition can arise, we will drop it. 
We use the shorthand notation $g(x,u) = [g_1(x,u),\dots, g_{n_g}(x,u)]^\top$.
Finally, $V_N: \hat x_i \to \mbb{R}$ denotes the optimal value function of \eqref{eq:original_ocp}.

As usual in NMPC and EMPC, the receding horizon solution to \eqref{eq:original_ocp} implies the following closed-loop dynamics
\begin{equation} \label{eq:sys_closed_loop}
\hat  x_{i+1} = f(\hat x_i, \mpcLaw(\hat x_i)), \quad \hat x_0 \in \mbb{X}_0, i \in \mbb{N},
\end{equation}
where the superscript $\hat\cdot$ distinguishes actual system variables from their predictions and the MPC feedback $\mpcLaw:\mbb{X}\to\mbb{U}$ is defined as usual:
\[
\mpcLaw(\hat x) \doteq u^\star_0(\hat x).
\]

\subsection{Dissipativity-based Stability Results}
Recall the following standard  definition: a function $\alpha:\mbb{R}_0^+ \to \mbb{R}_0^+$ is said to belong to class $\mcl{K}$, if it is continuous, strictly increasing, and $\alpha(0) = 0$.
We begin by recalling a dissipativity notion for OCPs. 
\begin{definition}[Strict dissipativity]\label{def:DI}~\\
\begin{enumerate}
\vspace*{-.5cm}
\item System \eqref{eq:original_ocp_dyn} is said to be \emph{dissipative with respect to the steady-state pair $ \begin{bmatrix}\bar x &\bar u\end{bmatrix}^\top \in\mbb{Z}$}, 
if there exists a non-negative function $ S:\mbb{X} \to \mbb{R}$ such that for all $ \begin{bmatrix}\bar x &\bar u\end{bmatrix}^\top \in\mbb{Z}$
\begin{subequations} \label{eq:DI}
\begin{equation}\label{eq:DI_non_str}
 S(f(x, u)) - S(x) \leq \ell(x, u) -  \ell(\bar x, \bar u).
\end{equation}
\item If, additionally, there exists $\alpha_\ell\in\mcl{K}$ such that
\begin{equation}  \label{eq:DI_str}
 S(f(x, u)) - S(x) \leq -\alpha_\ell\left(\left\|(x, u)-(\bar x,\bar u)\right\|\right) + \ell(x, u) - \ell(\bar x, \bar u).
\end{equation}
\end{subequations}
then  \eqref{eq:original_ocp_dyn} is said to be \emph{strictly dissipative with respect to $ \begin{bmatrix}\bar x &\bar u\end{bmatrix}^\top \in\mbb{Z}$}. 

\item If, for all $N\in\mbb{N}$ and all $x_0 \in \mbb{X}_0$, the dissipation inequalities \eqref{eq:DI} hold along any optimal pair of  \eqref{eq:original_ocp}, 
then \emph{OCP \eqref{eq:original_ocp} is said to be (strictly) dissipative with respect to $
(\bar x, \bar u)$}.   \eBox
\end{enumerate}
\end{definition}
It is worth to be remarked that $\ell$ in  \eqref{eq:DI} is the stage cost of \eqref{eq:original_ocp}. Moreover, we define the so-called \textit{supply rate} $s:\mbb{Z} \to \mbb{R}$ as
\[
s(x,u) \doteq \ell(x, u) - \ell(\bar x, \bar u), 
\]
while we denote $ S$ in \eqref{eq:DI} as a \textit{storage function}. We then see that \eqref{eq:DI} are indeed dissipation inequalities \cite{Moylan14a, Willems72a, Willems07a}.

Finally, we note that in the literature on EMPC different variants of the dissipation inequality \eqref{eq:DI_str} are considered; i.e., in the early works \cite{Angeli12a,Mueller14a,Gruene13a} strictness is required only in $x$, while more recent works consider strictness in $x$ and $u$, see \cite[Rem. 3.1]{kit:faulwasser18c} and \cite{epfl:faulwasser15g,epfl:faulwasser15h,Gruene16a}. 

Notice that the strict dissipation inequality \eqref{eq:DI_str} implies that $(\bar x, \bar u)$ is the  unique globally optimal solution to the following Steady-state Optimization Problem (SOP)
\begin{align}
\min_{ x,~  u} ~\ell(x,u) \quad
\text{subject to }\quad  x = f(x,u) \text{ and } \begin{bmatrix}x &u\end{bmatrix}^\top \in\mbb{Z}. \label{eq:SOP}
\end{align}
As we will assume strict dissipativity throughout this chapter, the unique globally optimal solution to \eqref{eq:SOP} is denoted by superscript $\bar\cdot$. Moreover, the optimal Lagrange multiplier vector of the equality constraint $ x = f(x,u) $ is denoted as $\bar\lambda \in\Rnx$ (assuming it is unique).

Subsequently, we compare three different  EMPC schemes (i)---(iii) based on \eqref{eq:original_ocp} which differ in terms of the terminal penalty $V_\mathrm{f}$ and the terminal constraint $\mbb{X}_\mathrm{f}$. These schemes are defined as follows:
\begin{itemize}
\item[(i)] ~~~$V_\mathrm{f}(x) = 0$ \quad \quad~ and \qquad $\mbb{X}_\mathrm{f} = \{\bar x\}$;
\item[(ii)]~~~$V_\mathrm{f}(x) = 0$ \quad \quad~ and \qquad $\mbb{X}_\mathrm{f} = \mbb{R}^{n_x}$;
\item[(iii)]~~~$V_\mathrm{f}(x) = \bar\lambda^\top x$ \quad \,and \qquad $\mbb{X}_\mathrm{f} = \mbb{R}^{n_x}$.
\end{itemize}
Note that in the third scheme the optimal Lagrange multiplier vector $\bar\lambda$ of \eqref{eq:SOP} is used in the terminal penalty definition. Moreover, we remark that we consider these three schemes, as their stability proofs rely on dissipativity of OCP \eqref{eq:original_ocp} while explicit knowledge of a storage function is not required. 

\subsubsection*{Asymptotic Stability via Terminal Constraints}
We begin our comparison by recalling conditions under which the EMPC scheme defined by  \eqref{eq:original_ocp} yields $\{$practical, asymptotic$\}$ stability  of the closed-loop system~\eqref{eq:sys_closed_loop}.
\begin{assumption}[Dissipativity]\label{ass:strDI}
OCP \eqref{eq:original_ocp} is strictly dissipative with respect to $ \begin{bmatrix}\bar x &\bar u\end{bmatrix}^\top \in \mbb{Z}$ in the sense of Definition \ref{def:DI} (iii). 
\end{assumption}
In \cite{Angeli12a} the following result analyzing the EMPC scheme \eqref{eq:original_ocp} for $V_\mathrm{f}(x) = 0$ and $\mbb{X}_\mathrm{f} = \{\bar x\}$ has been presented.

\begin{theorem}[Asymptotic stability of EMPC with terminal constraints] \label{thm:stab}~\\
Let Assumption \ref{ass:strDI} hold. Suppose that $V_\mathrm{f}(x) = 0$ and $\mbb{X}_\mathrm{f} = \{\bar x\}$ in \eqref{eq:original_ocp}. 
Moreover, suppose that $ S$ and $V_N$ are continuous at $x = \bar x$. 

Then, for all initial conditions $\hat x_0$ for which OCP \eqref{eq:original_ocp} is feasible, it remains feasible for $i\geq 0$, and the closed-loop system \eqref{eq:sys_closed_loop} is asymptotically stable at $\bar x$. 
\end{theorem}
This result, and its precursor in \cite{Diehl11a}, are appealing as no knowledge about the storage function $ S$ is required. However, knowledge about the optimal steady-state is required to formulate the terminal equality constraint $\mbb{X}_\mathrm{f} =\{\bar x\}$. Indeed, a dissipativity-based scheme with terminal inequality constraints, which requires knowledge of a storage function,  has been proposed  in \cite{Amrit11}. 

\subsubsection*{Practical Asymptotic Stability without Terminal Constraints and Penalties}
Consequently, and similarly to conventional tracking NMPC schemes, the extension to the case without terminal constraints ($\mbb{X}_\mathrm{f} = \R^{n_x}$) has been of considerable interest.\footnote{Note that the input-state constraints defined via $\mbb{Z}$ from \eqref{eq:Zset} are imposed also at the end of the prediction horizon, i.e., at $k =N$.} This has been done in \cite{Gruene13a,kit:faulwasser18c,Stieler14b} using further assumptions. To this end, we consider a set of initial conditions $ \mbb{X}_0 \subset \Rnx$.
\begin{assumption}[Reachability and local controllability]
\label{ass:Reach}~\\
\begin{enumerate}
\vspace*{-.5cm}
\item[(i)] For all $x_0 \in \mbb{X}_0$, there exists an infinite-horizon admissible input $u(\cdot;x_0)$, and constants $c \in (0,\,\infty)$, $\rho\in [0,1)$, such that
\[
\|(x(k;x_0, u(\cdot;x_0)), u(k;x_0)) - (\bar x, \bar u)\| \leq c\rho^k,
\]
i.e., the steady state $\bar x$ is exponentially reachable.
\item[(ii)] The Jacobian linearization $(A,B)$ of system \eqref{eq:original_ocp_dyn} at  $ \begin{bmatrix}\bar x &\bar u\end{bmatrix}^\top\in \inte\,{\mbb{Z}}$ is  $n_x$-step reachable.\footnote{We remark that $n_x$-step reachability of $x^+ = Ax +Bu$ implies that starting from $x=0$ one can reach any $x\in\Rnx$ within $n_x$ time steps; and one can steer any  $x \not=0$ to the origin within $n_x$ time steps, cf. \cite{Weiss72a}. In other words, $n_x$-step reachability implies $n_x$-step controllability.} 
\item[(iii)]The optimal steady state satisfies $ (\bar x, \bar u) \in \inte\,{\mbb{Z}}$.
\end{enumerate}
\end{assumption}
In \cite{Gruene13a,kit:faulwasser18c} the following result analyzing the EMPC scheme \eqref{eq:original_ocp} for $V_\mathrm{f}(x) = 0$ and $\mbb{X}_\mathrm{f} = \mbb{R}^{n_x}$ has been presented. It uses the notion of a class $\mcl{K}\mcl{L}$ function. A function $\gamma:\R_0^+\to \R_0^+$ is said to belong to class $\mcl{L}$ if continuous, strictly decreasing, and $\lim_{s\to\infty}\gamma(s) = 0$. A function $\beta: \R_0^+\times\R_0^+ \to \R_0^+$ is said to belong to class $\mcl{K}\mcl{L}$ if it is of class $\mcl{K}$ in its first argument and of class $\mcl{L}$ in its second argument.

\begin{theorem}[Practical stability of EMPC without terminal constraints] \label{thm:pstab}~\\
Let Assumptions \ref{ass:strDI} and \ref{ass:Reach} 
hold. Suppose that $\mbb{Z}$ is compact and that $V_\mathrm{f}(x) = 0$ and $\mbb{X}_\mathrm{f} = \mbb{R}^{n_x}$ in \eqref{eq:original_ocp}. 
Then the closed-loop system \eqref{eq:sys_closed_loop} has the following properties:
\begin{enumerate}
\item[(i)] If $x_0 \in \mbb{X}_0$ and $N \in \mbb{N}$ is sufficiently large, then OCP \eqref{eq:original_ocp} is feasible for all $i \in \N$.
\item[(ii)] There exist $\gamma(N) \in \R^+$ and $\beta \in \mcl{K}\mcl{L}$ such that, for all $x_0 \in \mbb{X}_0$, the closed-loop trajectories generated by  \eqref{eq:sys_closed_loop} satisfy
\[
\|x_i - \bar x\| \leq \max\{\beta(\|x_0 - \bar x\|,\,t),\, \gamma(N)\}. 
\]   
\end{enumerate}
\end{theorem}
The proof of this result is based on the fact that the dissipativity property of  OCP \eqref{eq:original_ocp} implies a turnpike property of the underlying OCP \eqref{eq:original_ocp}; it can be found in \cite{kit:faulwasser18c}. The original versions in \cite{Gruene17a, Gruene13a} do not entail the recursive feasibility statement. Under additional continuity assumptions on $S$ and on the optimal value function one can show that the size of the neighborhood---i.e. $\gamma(N)$ in (ii)---converges to $0$ as $N\to\infty$, cf. \cite[Lem. 4.1 and Thm. 4.1]{kit:faulwasser18c}. Similar results can also be established for the continuous-time case \cite{epfl:faulwasser15g}.

We remark that the size of the practically stabilized neighborhood of $\bar x$  might be quite large---especially for short horizons. We will illustrate this via an example in Section \ref{sec:Examples}; further ones can be found in \cite{kit:zanon18a,kit:faulwasser18c}. 
Moreover, comparison of Theorem \ref{thm:stab} and Theorem \ref{thm:pstab} reveals a gap: while with terminal constraints $\mbb{X}_\mathrm{f} = \{\bar x\}$, strict dissipativity implies asymptotic stability of the EMPC loop, without terminal constraints merely practical stability is attained. 

\subsubsection*{Asymptotic Stability via Gradient Correction}
We turn towards the question of how this gap can be closed. In \cite{kit:zanon18a,kit:faulwasser18e_2} we analyzed EMPC using $V_\mathrm{f}(x) = \bar\lambda^\top x$ and $\mbb{X}_\mathrm{f} = \mbb{R}^{n_x}$.

In order to state the core stability result concisely we first recall a specific linear-quadratic approximation of OCP \eqref{eq:original_ocp}. 
To this end and similar to  \cite{kit:faulwasser18e_2}, we consider the following Lagrangian\footnote{Arguably, one could denote $L$ from \eqref{eq:L} also as a \textit{Hamiltonian}. However, as we are working with discrete-time systems we stick to the nonlinear programming terminology. Readers familiar with optimal control theory will hopefully experience no difficulties to translate this to a continuous-time notions, cf. \cite{kit:zanon18a}.} of OCP \eqref{eq:original_ocp}
\begin{subequations} \label{eq:L}
	\begin{align}
	L_0 &\doteq \lambda_0^\top(x_0-\hat x_i), &  \quad k &= 0, \\
	L_k &\doteq \ell(x_k,u_k) +\lambda_{k+1}^\top(f(x_k, u_k) -x_{k+1})+\mu_k^\top g(x_k, u_k),& \quad k &\in   \mathbb{I}_{[0,N-1]}, \\
	L_N &\doteq V_\mathrm{f}(x_N) +\mu_N^\top g(x_N, u_N), &  \quad k &= N.
	\end{align}
\end{subequations}
	Accordingly we have the Lagrangian of SOP \eqref{eq:SOP} as
	\begin{equation}
	\bar L \doteq \ell(x,u) +\lambda^\top(f(x, u) -x)+\mu^\top g(x, u).
	\end{equation}

	In correspondence with this Lagrangian, we denote the optimal dual variables of SOP \eqref{eq:SOP} as $\bar\lambda, \bar\mu\geq 0$. The optimal adjoint and multiplier trajectories of OCP \eqref{eq:original_ocp} are written as $\lambda^\star(\cdot; \hat x_i)$ and $\mu^\star(\cdot; \hat x_i)\geq 0$.
Subsequently, we consider the following linear time invariant LQ-OCP
\begin{subequations}
	\label{eq:lqr_ocp}
	\begin{align}
	\min_{\boldsymbol{x}, \boldsymbol{u}} \ \ &  \sum_{k=0}^{N-1}
	\textstyle{\frac{1}{2}}\begin{bmatrix}x_k\\ u_k\end{bmatrix}^\top W\begin{bmatrix}x_k\\ u_k\end{bmatrix} + w^\top z_k +
\frac{1}{2}x_N^\top P_N x_N +  x_N^\top p_N  \label{eq:lqr_objective} \\
	\text{s.t.} \ \  
	& x_0 = \hat x_0,\\
	& x_{k+1} = Ax_k + Bu_k, \quad k  \in \mathbb{I}_{[0,N-1]}, \\
	& Cx_k + Du_k - g(\bar z^\star) \leq 0, \quad k  \in \mathbb{I}_{[0,N]}.
	\end{align}
	where the linear dynamics and path constraints are defined via the Jacobians $A = f_x$, $B =f_u$, $C = g_x$, $D =g_u$,
	and the quadratic objective is given by 
	 \[
	 W=\begin{bmatrix}Q\phantom{^\top} & S \\ S^\top & R\end{bmatrix}, \qquad w = \begin{bmatrix}q \\ r \end{bmatrix},
	 \]
	  with 
	  \begin{align*}
	  	Q = \bar L_{xx}, && S = \bar L_{xu}, && R = \bar L_{uu}, && q = \ell_x, && r = \ell_u, && P_N =V_{\mathrm{f}, xx}, && p_N = V_{\mathrm{f},x},
	  \end{align*}
	\end{subequations}
	where the functions and derivatives above are all evaluated at the primal-dual optimal solution of the SOP \eqref{eq:SOP}, i.e., at $\bar x=0, \bar u=0, \bar \lambda, \bar \mu=0$. Observe that $\bar\mu = 0$ corresponds to Assumption \ref{ass:Reach} (iii) and that $\bar x=0, \bar u=0$ is assumed without loss of generality. However, in general $\bar \lambda \neq 0$, as detailed in~\cite{kit:faulwasser18e_2,kit:zanon18a}.
	Moreover, we denote the optimal primal and dual variables of the LQ-OCP \eqref{eq:lqr_ocp} by
	\begin{align*}
\mbf{u}_{LQ}^\star(\hat x_i) &\doteq  \begin{bmatrix}u_{LQ, 0}^\star(\hat x_i) ^\top & \dots& u_{LQ,N-1}^\star(\hat x_i) ^\top\end{bmatrix}^\top, \\
\mbf{\xi}_{LQ}^\star(\hat x_i) &\doteq  \begin{bmatrix}\xi_{LQ,0}^\star(\hat x_i) ^\top & \dots& \xi_{LQ,N}^\star(\hat x_i) ^\top\end{bmatrix}^\top, &\quad \xi \in\{x, \lambda, \mu\}.
\end{align*}

	\begin{assumption}[Local approximation and stabilization properties]\label{ass:localProp}~\\
	\begin{itemize} \vspace*{-4mm}
	\item[(i)] ~~The SOP \eqref{eq:SOP} is such that the optimal dual variable $\bar\lambda$ is unique. 
	\item[(ii)] ~~There exists a horizon length $N<\infty$ such that the receding horizon feedback generated by the LQ-OCP \eqref{eq:lqr_ocp} stabilizes the linearized system $(A,~B)$ at some point $(\tilde x, \tilde u)$ which may differ from $(\bar x, \bar u)$.
	\item[(iii)] ~~The primal and dual optimal solutions of OCP \eqref{eq:original_ocp} and LQ-OCP \eqref{eq:lqr_ocp} satisfy
	\[
	\|\mbf{\xi}_{LQ}^\star(\hat x_i)-\mbf{\xi}^\star(\hat x_i)\| =O\left(\|\hat x_i\|^2\right), \qquad   \xi \in \{x, u,\lambda, \mu\}.
	\]
	\end{itemize}
	\end{assumption}
Part (i) of the above assumption is an implicit requirement of Linear Independence Constraint Qualification (LICQ) in \eqref{eq:SOP}, while Part (ii) is essential for our latter developments as it allows the assessment of asymptotic stability. Finally, Part (iii) can be read as a regularity property of the OCP \eqref{eq:original_ocp}. In \cite[Prop. 1]{kit:faulwasser18e_2} we have discussed that this can be enforced, for example, via strict complementarity. 

In \cite{kit:faulwasser18e_2} the following result analyzing the EMPC scheme \eqref{eq:original_ocp} for $V_\mathrm{f}(x) = \bar\lambda^\top x$ and 
$\mbb{X}_\mathrm{f} =  \mbb{R}^{n_x}$ has been presented. The continuous-time counterpart can be found in~\cite{kit:zanon18a}.

\begin{theorem}[Asymptotic stability of EMPC with linear terminal penalty] \label{thm:astab}~\\
Let Assumptions \ref{ass:strDI}, \ref{ass:Reach} and \ref{ass:localProp} hold. Suppose that $\mbb{Z}$ is compact and that $V_\mathrm{f}(x) = \bar\lambda^\top x$ and $\mbb{X}_\mathrm{f} = \mbb{R}^{n_x}$ in \eqref{eq:original_ocp}. 
Then, if $x_0 \in \mbb{X}_0$, the closed-loop system \eqref{eq:sys_closed_loop}, there exists a sufficiently large finite horizon $N\in\N$, such that:
\begin{enumerate}
\item[(i)] 
OCP \eqref{eq:original_ocp} is feasible for all $i \in \N$.
\item[(ii)] 
 $\bar x$ is an exponentially stable equilibrium of the closed-loop system \eqref{eq:sys_closed_loop}. 
\end{enumerate}
\end{theorem}
Leaving the technicalities of Assumption \ref{ass:localProp} aside, it is fair to ask what inner mechanisms yield that the linear end penalty $V_\mathrm{f}(x) = \bar\lambda^\top x$ makes the difference between practical and asymptotic stability? Moreover, a comparison of the results of Theorems \ref{thm:stab}, \ref{thm:pstab} and \ref{thm:astab} is clearly in order. We will comment on both aspects below.

\section{Comparison}  \label{sec:Compare}
We begin our comparison of the three EMPC schemes with an observation: neither Theorems \ref{thm:stab} nor Theorem \ref{thm:pstab} use any sort of statement on dual variables---i.e., Lagrange multipliers $\mu$ and adjoints $\lambda$. Indeed, a close inspection of the proofs of Theorem \ref{thm:stab}~\cite{Angeli12a} and of Theorem \ref{thm:pstab}~\cite{Gruene13a,kit:faulwasser18c}, confirms that they are solely based on primal variables. Actually, despite the crucial nature of optimization for NMPC~\cite{Gruene17a,Rawlings17}, the vast majority of NMPC proofs does not involve any information on dual variables. However, as documented in \cite{kit:faulwasser18c,kit:zanon18a} the proof of Theorem \ref{thm:astab} relies heavily on dual variables.

\subsection{Discrete-time Euler-Lagrange Equations}
Based on this observation, we begin our comparison of the three EMPC schemes, by detailing the first-order Necessary Conditions of Optimality (NCO), i.e., the KKT conditions of OCP \eqref{eq:original_ocp} which we present in the form of discrete-time Euler-Lagrange equations~\cite{Bryson99a}.

The overall Lagrangian of OCP \eqref{eq:original_ocp} reads as
\[
\mcl{L} \doteq \sum_{k=0}^N L_k,
\]
with $L_k$ from \eqref{eq:L}. The first-order NCO are given by $\nabla \mcl{L}=0$, which entails 
\begin{subequations} \label{eq:NCO}
\begin{align}
\mcl{L}_\lambda = 0& \quad \Rightarrow& x_{k+1} &= f(x_k, u_k), && x_0 = \hat x_i, \\
\mcl{L}_x = 0& \quad\Rightarrow& \lambda_{k} &=  f_x^\top\lambda_{k+1} +\ell_x+ g_x^\top\mu,\\
\mcl{L}_u = 0& \quad\Rightarrow& 0 &= f_u^\top\lambda_k+\ell_u +g_u^\top\mu.
\end{align}
The reader familiar with the Euler-Lagrange equations might have noticed that \eqref{eq:NCO} misses a crucial piece, i.e., the terminal constraints on either the primal state $x_N$ or the dual (adjoint) variable $\lambda_N$. Depending on the specific choice for $V_\mathrm{f}$ and $\mbb{X}_\mathrm{f}$, and assuming that the constraint \eqref{eq:original_ocp_pc} is not active at $k=N$---which implies $\mu_N = 0$---, these terminal constraints read:
\begin{equation} \label{eq:NCO_bnd}
\begin{array}{lllll}
\text{(i)\phantom{ii} }~ x_N = \bar x,&\lambda_N\in\mathbb{R}^{n_x},~ &\text{ if }  V_\mathrm{f}(x_N) = 0 &\text{ and }~ \mbb{X}_\mathrm{f} = \bar x; \\
\text{(ii)\phantom{i} }~x_N\in\mathbb{R}^{n_x},&\lambda_N = 0,~ &\text{ if } V_\mathrm{f}(x_N) = 0 &\text{ and }~ \mbb{X}_\mathrm{f} = \mbb{R}^{n_x}; \\ 
\text{(iii) }~x_N\in\mathbb{R}^{n_x},&\lambda_N = \bar\lambda,~ &\text{ if }  V_\mathrm{f}(x_N) = \bar\lambda^\top x_N &\text{ and }~ \mbb{X}_\mathrm{f} = \mbb{R}^{n_x}. 
\end{array}
\end{equation}
\end{subequations}
Before proceeding, we remark that the full KKT conditions would also comprise primal feasibility, dual feasibility and complementarity constraints. We do not detail those here, as  the discrete-time Euler-Lagrange equations \eqref{eq:NCO} provide sufficient structure for our analysis, and all left-out optimality conditions coincide for the three problem formulations.

Comparing the three EMPC schemes (i)--(iii) at hand, the first insight is obtained from \eqref{eq:NCO_bnd}: the only difference in the optimality conditions is the boundary constraint. We comment on the implications of this fact next.

\subsection{Primal Feasibility and Boundary Conditions of the NCO}

In case of Scheme (i), the NCO comprise the primal terminal constraint (primal boundary condition) $ x_N = \bar x$. In fact, for any $\hat x_0 \in \mbb{X}_0$ finite-time reachability of $\bar x$ must be given in order for \eqref{eq:original_ocp} to be feasible. In other words, \textit{feasibility of $x_N = \bar x$ is a necessary condition for OCP \eqref{eq:original_ocp} to admit optimal solutions}.

In case of  Schemes (ii) and (iii), the NCO \eqref{eq:NCO} comprise dual boundary conditions $\lambda_N = 0$, respectively, $\lambda_N = \bar\lambda$. The crucial difference between primal and dual boundary conditions is that the existence of an optimal solution certifies that the latter are feasible. In terms of logical implications---and provided that OCP \eqref{eq:original_ocp} viewed as an NLP satisfies LICQ and the assumption on continuity of problem data---we have that
\begin{multline*}
\text{primal feasibility of OCP \eqref{eq:original_ocp}} \quad \Rightarrow\\
\quad \text{existence of optimal solutions in OCP \eqref{eq:original_ocp}} \quad \Rightarrow\\
\quad \text{dual feasibility of NCO \eqref{eq:NCO}}.
\end{multline*}
Let
\[
\mbb{F}_{\mbb{X}_\mathrm{f}}^{V_\mathrm{f}}(x_0,N) \subseteq \mbb{Z}\times \dots \times\mbb{Z}\subseteq\mbb{R}^{N\cdot(n_u+n_x+1)}
\]
denote the feasible set of OCP \eqref{eq:original_ocp} parametrized by the initial condition $x_0$ and the horizon length $N$, where the subscript $\cdot_{\mbb{X}_\mathrm{f}}$ refers to the considered terminal constraint and the superscript $\cdot^{V_\mathrm{f}}$ highlights the terminal penalty. 
In terms of feasible sets, the differences between the three considered schemes and thus the implications of their respective primal and dual boundary constraints can be expressed as follows
\[
\mbb{F}_{\bar x}^0(x_0, N) \subseteq \mbb{F}_{\mbb{R}^{n_x}}^0(x_0,N) \equiv \mbb{F}_{\mbb{R}^{n_x}}^{\bar\lambda^\top x}(x_0, N), \qquad \forall \, x_0 \in \mbb{X}_0, \forall \, N \in \mbb{N}.
\]
The first set relation follows from the fact that any feasible solution of Scheme (i) is also feasible in Schemes (ii) and (iii), but the opposite is not true. The second set relation is also evident, since Schemes (ii) and (iii) differ only in the objective function and not in the primal constraints.

\subsection{Invariance of $\bar x$ under EMPC}
The next aspect we address in terms of comparing  Schemes (i) -- (iii) is related to the invariance of $\bar x$ under the EMPC feedback, which, as we shall see, also crucially depends on the boundary constraints \eqref{eq:NCO_bnd}.

\begin{figure}[h]
	\begin{center}
		\includegraphics[angle=0,origin=c,keepaspectratio,width=0.9\textwidth]{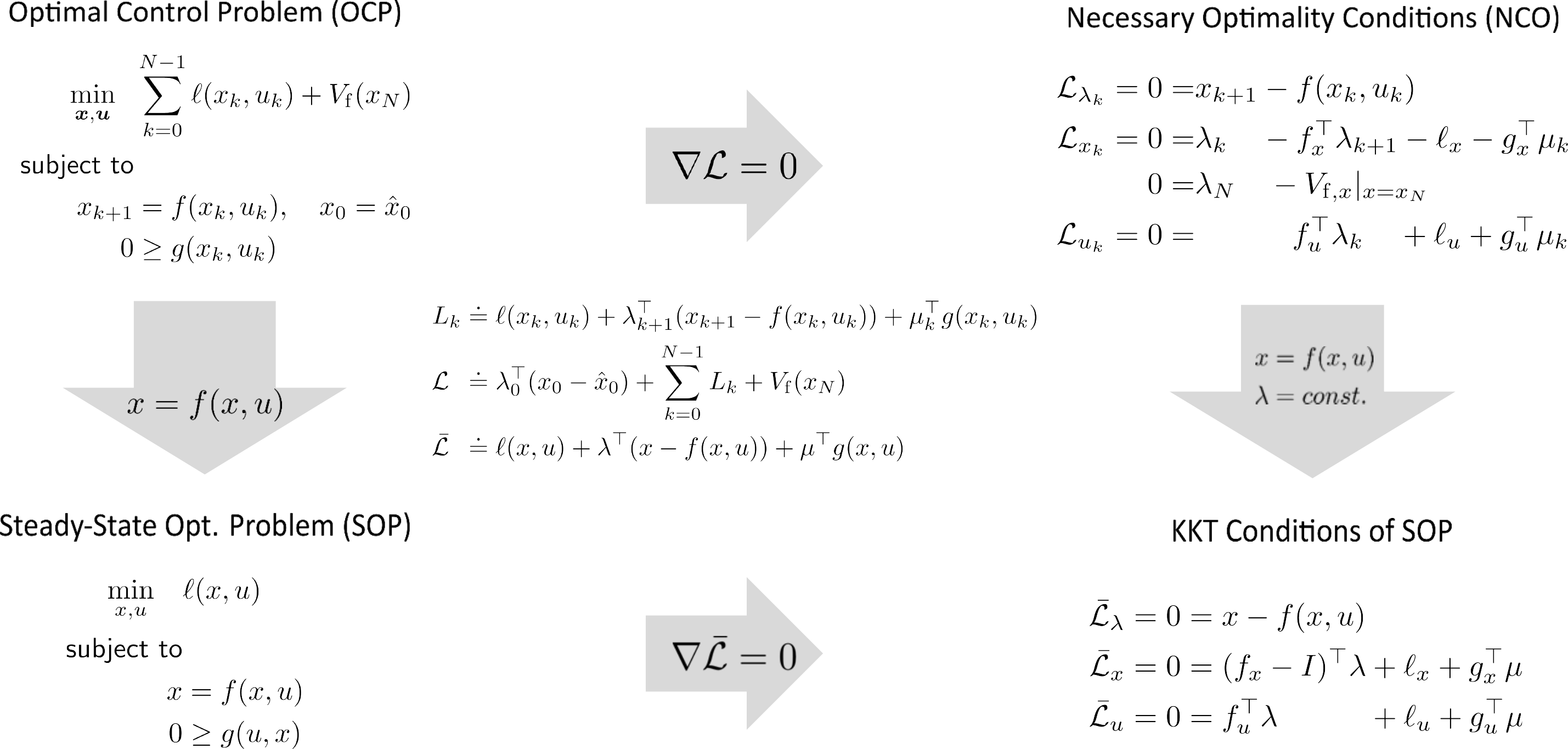}
	\end{center}
	\caption{Relation of OCP and the corresponding Euler-Lagrange Equations with the SOP and the corresponding KKT conditions.}
	\label{fig:equivalence}
\end{figure}

The invariance of $\bar x$ under the EMPC feedback can be  analyzed turning to the NCO \eqref{eq:NCO} which entail
\begin{align*}
x_{k+1} &= f(x_k, u_k),\\
\lambda_{k} &=  f_x^\top\lambda_{k+1} +\ell_x+ g_x^\top\mu,\\
0 &= f_u^\top\lambda_k+\ell_u +g_u^\top\mu.
\end{align*}
Recall that the SOP \eqref{eq:SOP} implies the (partial) KKT optimality conditions
\begin{subequations} \label{eq:KKT}
\begin{align}
\bar x &= f(\bar x, \bar u), \\
\bar \lambda &=  f_x^\top\bar\lambda +\ell_x+ g_x^\top\bar\mu,\\
0 &= f_u^\top\bar\lambda+\ell_u +g_u^\top\bar \mu,
\end{align}
\end{subequations}
for which due to Assumption \ref{ass:strDI} $\bar x, \bar u, \bar \lambda, \bar \mu =0$ is the unique solution. Note that 
$\bar\mu = 0$ follows from $ (\bar x, \bar u) \in \inte\,{\mbb{Z}}$. In other words, the KKT conditions of SOP \eqref{eq:SOP} coincide with the steady state version of the NCO \eqref{eq:NCO}. As we will see this observation, which is also sketched in Figure \ref{fig:equivalence}, is crucial in analyzing invariance  of $\bar x$ under the EMPC feedback. We mention that, to the best of our knowledge, the first usages of this observation appear to be~\cite{Trelat15a,kit:zanon18a}. 

Invariance  of $\bar x$ means
\begin{equation} \label{eq:invarianceEMPC}
\bar x = f(\bar x, \mpcLaw_{\mathbb{X}_f}^{V_\mathrm{f}}(\bar x)).
\end{equation}
This invariance holds if and only if, for the considered EMPC scheme
\[
\mpcLaw_{\mathbb{X}_f}^{V_\mathrm{f}}(\bar x) = \bar u.
\]
Here, instead of providing fully detailed proofs, we focus on the crucial system-theoretic aspects. 

Assumption \ref{ass:localProp} suggests to consider the NCO of the LQ approximation \eqref{eq:lqr_ocp}. Taking $ (\bar x, \bar u) \in \inte\,{\mbb{Z}}$ into account those NCO entail
\begin{align*}
\bar x &= A\bar x +B\bar u,\\
\lambda_{k} &=  A^\top\lambda_{k+1} +Q \bar x + q,\\
0 &= B^\top\lambda_k+R\bar u + r.
\end{align*}
Assuming that $\det R \not= 0$ and neglecting the first equation, we obtain
\begin{subequations} \label{eq:LQNCO}
\begin{align}
\lambda_{k} &=  \phantom{-}A^\top\lambda_{k+1} +Q \bar x + q, \label{eq:dynlambda}\\
\bar u &= -R^{-1}\left(B^\top\lambda_k+ r\right). \label{eq:outlambda}
\end{align}
\end{subequations}
This in turn can be read as a linear uncontrolled system \eqref{eq:dynlambda} with a linear output equation  \eqref{eq:outlambda}. 
Importantly, uniqueness of the steady state adjoint $\bar\lambda$ and controllability of $(A, B)$ ($\Leftrightarrow$ observability of $(A^\top, B^\top)$) imply that $\lambda_k = \lambda_{k+1} = \bar \lambda$ is the only solution \eqref{eq:LQNCO}, see~\cite{kit:faulwasser18e_2}. This shows that the closed-loop invariance condition \eqref{eq:invarianceEMPC} holds at $\bar x$ if and only if the optimal adjoint satisfies $\lambda^\star_0(\bar x) = \bar \lambda$, i.e., if the state initial condition is $\bar x$ then that adjoint at time $0$ is $ \bar \lambda$. 

Now, consider the boundary conditions of the adjoint $\lambda$ as given in \eqref{eq:NCO_bnd}. We start with case (ii), i.e., $ V_\mathrm{f}(x_N) = 0$, $\mbb{X}_\mathrm{f} = \mbb{R}^{n_x}$ and have $\lambda_N = 0$. This implies
\[
\lambda_{k} =  A^\top\lambda_{k+1} +Q \bar x + q, \qquad \lambda_N = 0.
\]
If $\lambda_0 = \bar\lambda$ would hold then the linearity of the adjoint dynamics implies for all $k\in \mbb{I}_{[0,N]}$ that $\lambda_k\equiv \bar \lambda$. This, however, contradicts the boundary constraint $\lambda_N = 0$. Hence the adjoint has to leave the optimal steady state value directly in the first time-step.\footnote{Likewise, considering the adjoint dynamics backwards in time from $k=N$ to $k=0$ implies that $\bar\lambda$ is never reached due to linearity of the dynamics.
}
A formal proof of the above considerations can be found in~\cite{kit:faulwasser18e_2}. Moreover, we remark without further elaboration that in the continuous-time case, the mismatch between dual terminal constraint at $\lambda_N=0$ leads to a small-scale periodic orbit which appears in closed loop, cf. \cite{kit:zanon18a}. Finally, in case of singular OCPs with $\det \bar L_{uu} = \det R = 0$ the NCOs do not directly allow to infer the controls. One can arrive at one of two cases:
\begin{itemize}
\item For sufficiently long horizons, the open-loop optimal solution reach the optimal steady state $(\bar x, \bar u)$ \textit{exactly}. In this case, one says that the turnpike at $(\bar x, \bar u)$ is exact. Moreover, one can show even with $\lambda_N=0$ (i.e. no terminal penalty) invariance of $\bar x$ under the EMPC feedback law and thus also asymptotic stability is attained. We refer to~\cite{kit:faulwasser17a} for a detailed analysis of the continuous-time case.
\item Despite the OCP being singular, for all finite horizons, the open-loop optimal solution get only close to the optimal steady state $(\bar x, \bar u)$, i.e., the turnpike is not exact but \textit{approximate}~\cite{epfl:faulwasser15g}. Without any terminal penalty or primal terminal constraint, one cannot expect to achieve asymptotic stability in this case.
\end{itemize}

In case (iii), i.e., $ V_\mathrm{f}(x_N) = \bar\lambda^\top x_N$, $\mbb{X}_\mathrm{f} = \mbb{R}^{n_x}$ we arrive at 
\begin{align*}
\lambda_{k} &=  \phantom{-}A^\top\lambda_{k+1} +Q \bar x + q, \quad \lambda_N = \bar\lambda,\\
\bar u &= -R^{-1}\left(B^\top\lambda_k+ r\right). 
\end{align*}
In this case, the boundary condition $\lambda_N = \bar\lambda$ prevents the adjoint from leaving $\bar\lambda$.

Finally, for scheme (i), i.e., $ V_\mathrm{f}(x_N) = 0$, $\mbb{X}_\mathrm{f} = \{\bar x\}$, there is no boundary constraint for the adjoint.
Here invariance of $\bar x$ is directly enforced by the primal terminal constraint $x_N = \bar x$. Slightly simplifying, one may state that the local curvature implied by the strict dissipation inequality \eqref{eq:DI_str} makes staying at $\bar x$ cheaper than leaving and returning. Moreover, one can also draw upon the stability result from Theorem \ref{thm:stab} to conclude that the invariance condition \eqref{eq:invarianceEMPC} holds, which in our setting implies that the adjoint is also at steady state $\bar \lambda$.

\

\subsection{Bounds on the Stabilizing Horizon Length}
The next aspect we are interested in analyzing is the minimal stabilizing horizon length induced by the differences in Schemes (i) -- (iii). 

To this end, recall Assumption \ref{ass:Reach} (i), which requires exponential reachability of $\bar x$ from all $x_0 \in \mbb{X}_0$. 
This  also means that the set $\mbb{X}_0$ is controlled forward invariant. Put differently, for all $x_0 \in \mbb{X}_0$, there exist infinite-horizon controls such that the solutions stay in $\mbb{X}_0$ for all times.

Similarly to the feasible sets $\mbb{F}_{\bar x}^0(N, x)$, $\mbb{F}_{\mbb{R}^{n_x}}^0(N, x)$ and $\mbb{F}_{\mbb{R}^{n_x}}^{\bar \lambda^\top x}(N, x)$ we use super- and sub-scripts to distinguish the EMPC feedback laws $\mpcLaw_{\bar x}^0(N, x)$, $\mpcLaw_{\mbb{R}^{n_x}}^0(N, x)$ and $\mpcLaw_{\mbb{R}^{n_x}}^{\bar \lambda^\top x}(N, x)$ for the three EMPC schemes. Moreover, using the arguments $N, x$ we highlight the dependence of the feedback on the horizon length $N$.

In case of Scheme (i), the computation of the minimal stabilizing horizon length can be formalized via the folllowing bi-level optimization problem
\begin{subequations}\label{eq:Nstar_stab}
\begin{align} 
N^0_{\bar x} = \min_{N\in\mbb{N}}& ~N \\
\text{ subject to }& \quad \mbb{F}_{\bar x}^0(N, x) \not= \emptyset, &\quad \forall \, x &\in \mbb{X}_0, \label{eq:Nstar_stab_feas}\\
&\quad f(x, \mpcLaw_{\bar x}^0(N,x)) \in \mbb{X}_0, &\quad \forall \, x &\in \mbb{X}_0. \label{eq:Nstar_stab_invariance}
\end{align}
\end{subequations}
The first constraint  \eqref{eq:Nstar_stab_feas} encodes the observation that for Scheme (i) feasibility implies closed-loop stability of $\bar x$. The second constraint \eqref{eq:Nstar_stab_invariance} encodes that for any NMPC scheme to be stabilizing the set $\mbb{X}_0$ is indeed rendered forward invariant by the NMPC feedback.
The latter constraint implies a bi-level optimization nature of the problem: in order to solve \eqref{eq:Nstar_pstab} / \eqref{eq:Nstar_astab}, one needs to simulate the closed EMPC loop to obtain the feedback in \eqref{eq:Nstar_stab_invariance}.

Note that this constraint will not be (strongly) active in case of a terminal point constraint. This can be seen from the fact that leaving  \eqref{eq:Nstar_stab_invariance} out will not change the value of $N^0_{\bar x} $. The reason is that, provided the problem was feasible at the previous time step, the point-wise terminal constraint makes it immediate to construct a feasible guess for the NMPC problem~\eqref{eq:original_ocp}. Therefore, \eqref{eq:Nstar_stab} can be rewritten equivalently as a single-level optimization problem.
Here we include this constraint to simplify a comparison of the three NMPC schemes. 

 Notice that the problem above, if solved to global optimality, provides the true minimal stabilizing horizon length for Scheme (i). Yet, solving it is complicated by the fact that $\mbb{F}_{\bar x}^0(N, x_0) \not= \emptyset, \ \forall \, x_0 \in \mbb{X}_0$ is an infinite dimensional constraint. 
 Moreover, as long as one does not tighten exponential reachablity to finite-time reachability of $\bar x$ from all $x_0 \in \mbb{X}_0$, the optimal solution to \eqref{eq:Nstar_stab} might be $N^0_{\bar x} = \infty$.

The straightforward counterparts to \eqref{eq:Nstar_stab} for Schemes (ii) and (iii) read
\begin{subequations}\label{eq:Nstar_pstab}
\begin{align} 
N_{\mbb{R}^{n_x}}^0 = \min_{N\in\mbb{N}}& ~N \\
\text{ subject to }& \quad \mbb{F_{\mbb{R}^{n_x}}^0}(N, x) \not= \emptyset, &\quad \forall \, x &\in \mbb{X}_0, \label{eq:Nstar_pstab_feas}\\
&\quad f(x, \mpcLaw_{\mbb{R}^{n_x}}^0(N,x)) \in \mbb{X}_0, &\quad \forall \, x &\in \mbb{X}_0, \label{eq:Nstar_pstab_invariance}
\end{align}
\end{subequations}
respectively, 
\begin{subequations}\label{eq:Nstar_astab}
\begin{align} 
N_{\mbb{R}^{n_x}}^{\bar \lambda^\top x} = \min_{N\in\mbb{N}}& ~N \\
\text{ subject to }& \quad \mbb{F}_{\mbb{R}^{n_x}}^{\bar \lambda^\top x}(N, x) \not= \emptyset, &\quad \forall \, x &\in \mbb{X}_0, \label{eq:Nstar_astab_feas}\\
&\quad f(x, \mpcLaw_{\mbb{R}^{n_x}}^{\bar \lambda^\top x}(N,x)) \in \mbb{X}_0, &\quad \forall \, x &\in \mbb{X}_0. \label{eq:Nstar_astab_invariance}
\end{align}
\end{subequations}
If, in all problems \eqref{eq:Nstar_stab}--\eqref{eq:Nstar_astab}, the forward invariance constraints  \eqref{eq:Nstar_stab_invariance}--\eqref{eq:Nstar_astab_invariance} are inactive---which is indeed the case for \eqref{eq:Nstar_stab_invariance} or if there are no state constraints implied by $\mbb{Z}$---, the following relation is easily derived:
\[
N_{\mbb{R}^{n_x}}^0= N_{\mbb{R}^{n_x}}^{\bar \lambda^\top x}\leq N^0_{\bar x}.
\]
At first sight, this relation appears to be a rigorous advantage of Schemes (ii) and (iii).
However, for those schemes there is, to the best of our knowledge, no general guarantee that recursive feasibility alone implies asymptotic properties. As a matter of fact, Scheme (ii) only admits practical asymptotic stability properties. Hence for  Schemes (ii) and (iii) the horizon length computed via \eqref{eq:Nstar_pstab} or \eqref{eq:Nstar_astab} constitutes a lower bound on the minimal stabilizing horizon length of the underlying schemes. 
Finally, we remark that while for \eqref{eq:Nstar_stab} the invariance constraint \eqref{eq:Nstar_stab_invariance} is inactive, in
\eqref{eq:Nstar_pstab} and \eqref{eq:Nstar_astab}  the feasibility set constraints \eqref{eq:Nstar_pstab_feas} and \eqref{eq:Nstar_astab_feas} are inactive. The reason is that Assumption \ref{ass:Reach}(i) implies that $\mbb{X}_0$ is a controlled forward invariant set. In turn, this directly implies that for all $x\in \mbb{X}_0$ and any $N\in \mbb{N}^+$ the feasibility 
sets in \eqref{eq:Nstar_pstab_feas} and \eqref{eq:Nstar_astab_feas} are non-empty as there are no primal terminal constraints in the  underlying OCPs. 
However, inactivity of these constraints does not provide a handle to overcome the bi-level optimization nature of \eqref{eq:Nstar_pstab} and \eqref{eq:Nstar_astab}. Hence we turn towards an approximation procedure. 

\subsubsection*{Computational Approximation}
To approximate the solution to \eqref{eq:Nstar_stab}, we fix a set of initial conditions 
\[
\tilde{\mbb{X}}_0 \doteq \{x_0^j, \quad j =1, \dots, M \}.
\]
For all  samples $x_0^j$, we solve the minimum-time problem 
\begin{subequations}
	\label{eq:minN_ocp}
	\begin{align}
	N_{\bar x}^0(x_0^j) \doteq\min_{\boldsymbol{x}, \boldsymbol{u}, N} \ \ & N \hspace{-3em} \label{eq:minN_ocp_objective} \\
	\text{subject to} \ \  & x_0 = x_0^j, \label{eq:minN_ocp_jc}\\
	& x_{k+1} = f(x_k,u_k), &&\hspace{-0.5em} k\in \mathbb{I}_{[0,N-1]}, \label{eq:minN_ocp_dyn}\\
	& g(x_k,u_k) \leq 0, &&\hspace{-0.5em}  k\in \mathbb{I}_{[0,N]}, \label{eq:minN_ocp_pc}\\
	& x_N =\bar x. \label{eq:minN_ocp_tc}
	\end{align}
\end{subequations}
By virtue of Bellman's optimality principle, the minimum-time problem defined above allows one to conclude the behaviour of the closed-loop with primal terminal constraint. 
Whenever no solution to this problem is found, we define $N_{\bar x}^0(x_0^j) \doteq \infty$. Eventually, an approximation of \eqref{eq:Nstar_stab} is given by
\[
 N_{\bar x}^0\approx \max_{x_0^j \in \tilde{\mbb{X}}_0 } ~N_{\bar x}^0(x_0^j),
\]
provided that the samples $x_0^j$ are sufficiently dense and cover a sufficiently large subset of the x-projection of $\mbb{Z}$.
Observe that additional information is contained in the tuples $(x_0^j, N_{\bar x }(x_0^j))$: large values of $ N_{\bar x }(x_0^j)$ indicate that reaching $\bar x$ from $x_0^j$ is difficult. 

The conceptual counterpart to \eqref{eq:minN_ocp} for Schemes (ii) and (iii) is a forward simulation of the closed loop with $\dagger=\{0,\bar \lambda^\top x\}$ and
\begin{subequations}
\begin{align}
	N_{\mbb{R}^{n_x}}^\dagger(x_0^j) \doteq \min_{N} &~~N   \hspace{-3em}\\
\text{subject to} \ \   &x_0 = x_0^j, \\
 &x_{k+1} = f(x_k,\mpcLaw_{\mbb{R}^{n_x}}^\dagger(N,x_k)), &&\hspace{-0.5em} k\in \mathbb{I}_{[0,N_\mathrm{cl}]},\\
 &x_{N_\mathrm{cl}} \in \mathcal{B}_\rho(\bar x),
\end{align}
\end{subequations}
where $\mathcal{B}_\rho$ denotes a ball of radius $\rho$ and $N_\mathrm{cl}$ denotes the closed-loop simulation horizon which should ideally be infinite, but can in practice only be finite.
Obviously, the above problem requires one to solve a large number of OCPs when simulating the closed-loop. Moreover, one has to rigorously define the radius $\rho$, i.e., the accuracy by which the optimal steady-state should be attained. 
This approximation procedure leads to 
\[
N_{\mbb{R}^{n_x}}^\dagger\approx \max_{x_0^j \in \tilde{\mbb{X}}_0 } ~N_{\mbb{R}^{n_x}}^\dagger(x_0^j),\qquad \dagger=\{0,\bar \lambda^\top x\}.
\]

\section{Simulation Example} \label{sec:Examples}

We consider a system with state $x=\begin{bmatrix}
x_\mathrm{A} & x_\mathrm{B}
\end{bmatrix}^\top \in [0,1]^2$, control $u\in [0,20]$, stage cost and dynamics
\begin{subequations}
	\label{eq:ex_nonlin}
	\begin{align}
	\ell(x,u) &= -2ux_\mathrm{B} + 0.5 u + 0.1(u-12)^2, \\
	f(x,u) &= \begin{bmatrix}
	 x_\mathrm{A} + 0.01 \,u \,(1 - x_\mathrm{A}) - 0.12\, x_\mathrm{A} \\  x_\mathrm{B} + 0.01\, u \, x_\mathrm{B} + 0.12\, x_\mathrm{A} 
	\end{bmatrix}.
	\end{align}
\end{subequations}
This example has also been considered in \cite{kit:faulwasser18e_2}.
The optimal steady state is $\bar x^\star = [0.5, 0.5]^\top$, $\bar u^\star=12$ with $\bar \lambda^\star=[-100,-200]^\top$.

\begin{figure}
	\includegraphics[width=0.5\linewidth]{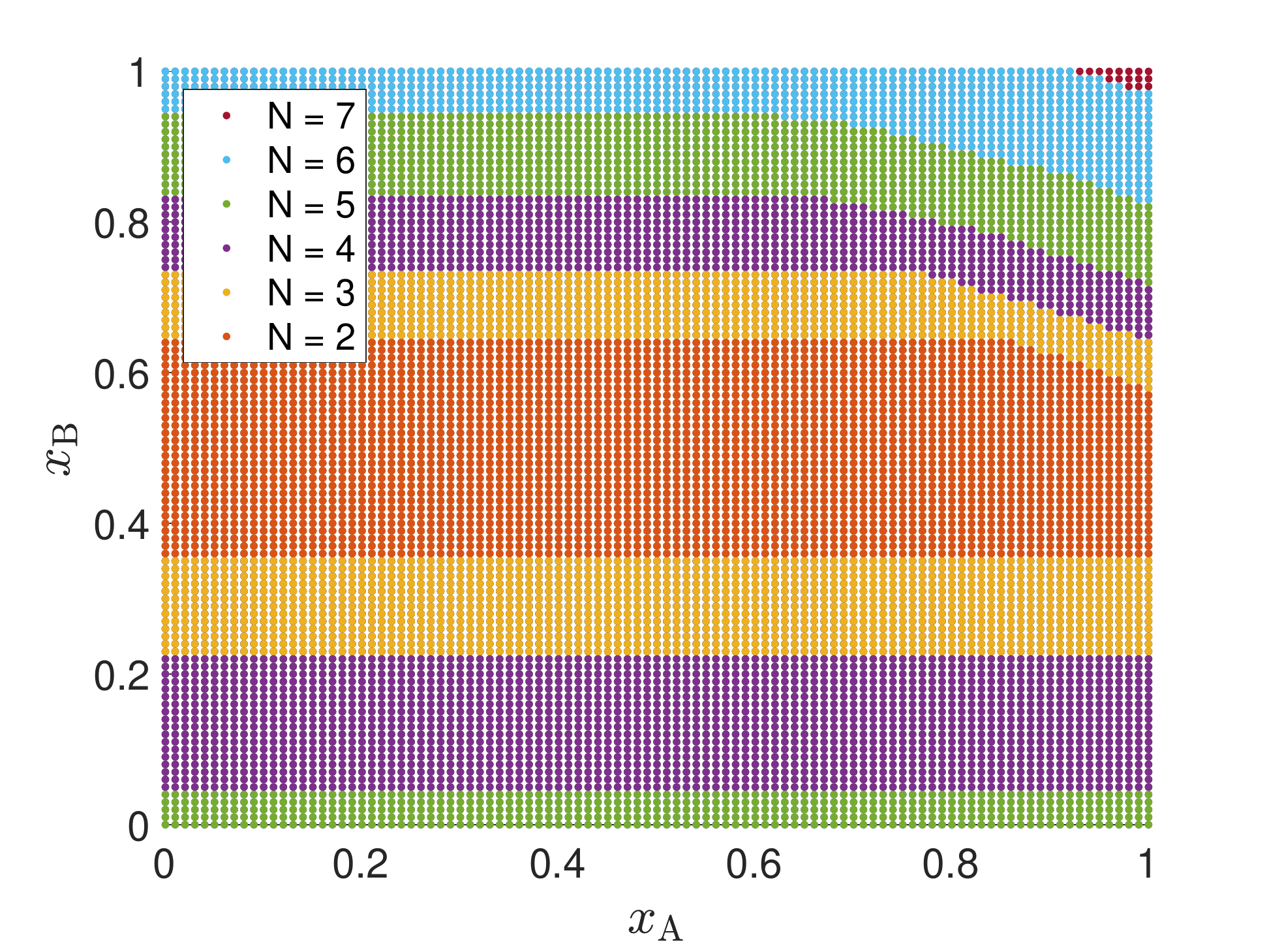}
	\includegraphics[width=0.5\linewidth]{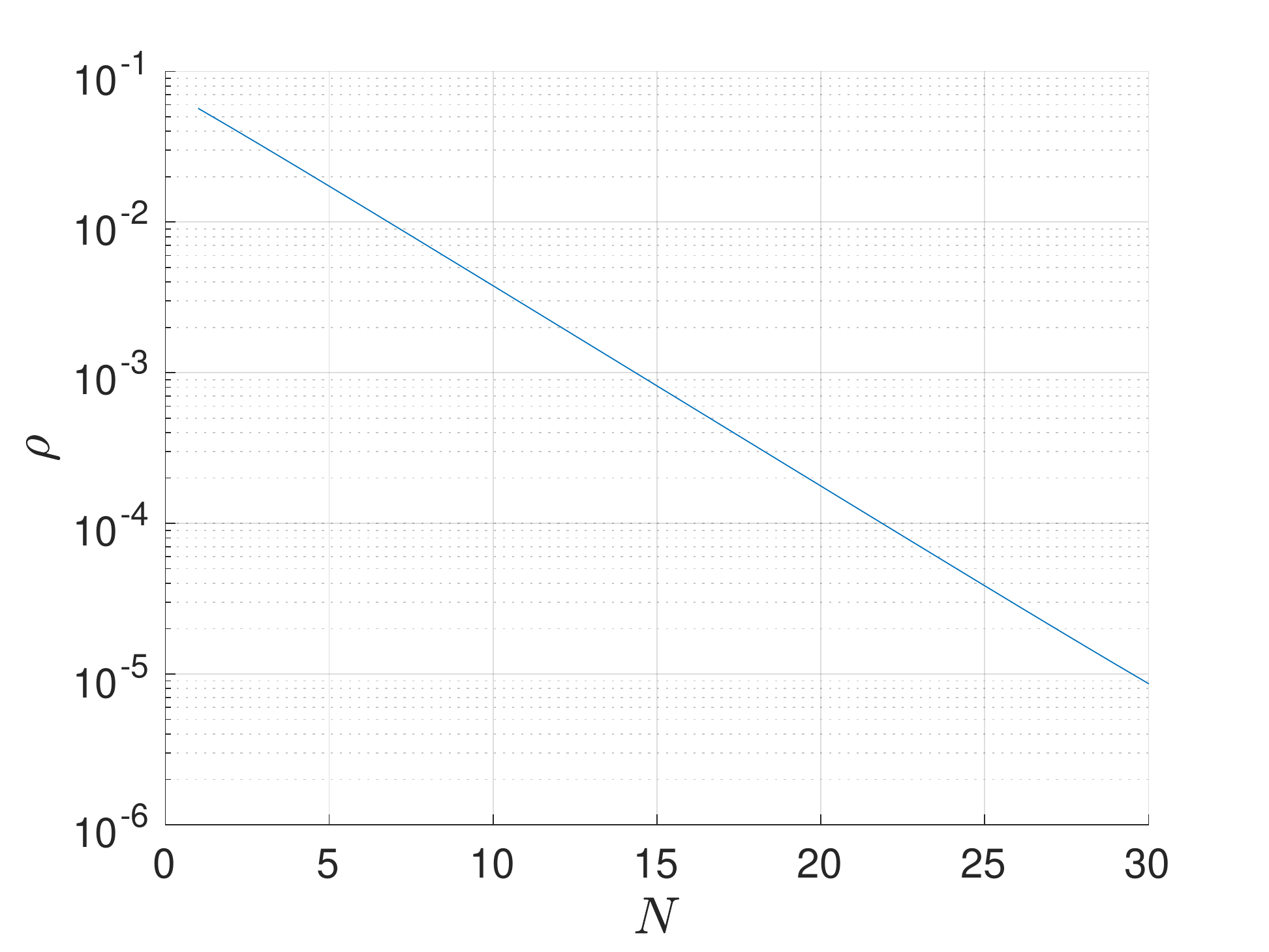}
	\caption{ Stabilizing horizon length. Left graph: $N_{\bar x}^0$ as a function of the initial state. Right graph: $\rho$ as a function of $N$ for scheme (ii). }
	\label{fig:stability}
\end{figure}

We computed the stabilizing horizon length for the three economic NMPC variants and observed that for the two schemes without terminal constraints in case $V_\mathrm{f}(x) = \bar \lambda^\top x$ we obtain stability with $N=1$ for any $\rho>0$, while for $V_\mathrm{f}(x) = 0$ the stabilizing horizon length is independent of the initial condition but depends on $\rho$ as displayed in Figure~\ref{fig:stability} (right). For the formulation with the terminal constraint $x_N = \bar x$, the stabilizing horizon length depends on the initial condition, as some initial conditions are potentially infeasible. We display the stability regions for several choices of $N$ in Figure~\ref{fig:stability} (left) and observe that for $N\geq7$ stability is obtained for all feasible initial states. These results are also summarized in Table \ref{tab:results}.
\begin{table}
\caption{Numerical approximation results of the minimal stabilizing horizon length for \eqref{eq:ex_nonlin}.}\label{tab:results}
\begin{center}
\begin{tabular}{l c || c | c}
EMPC& & Minimal stabilizing & Eventual deviation  \\
&& horizon length  for $x_0 \in \tilde{\mbb{X}}_0$& from $\bar x$\\
\hline
(i) $V_\mathrm{f}(x) = 0$ & $\mbb{X}_\mathrm{f} = \{\bar x\}$  & $N= 7$ & $0$ \\
(ii) $V_\mathrm{f}(x) = 0$ &  $\mbb{X}_\mathrm{f} = \mbb{R}^{n_x}$ & $N\geq 1$ & $<7.6\cdot 10^{-2} \ e^{-0.29N}$\\
(iii) $V_\mathrm{f}(x) = \bar\lambda^\top x$ & ~$\mbb{X}_\mathrm{f} = \mbb{R}^{n_x}$ & $N =1$ & $0$
\end{tabular}
\end{center}
\end{table}

\section{Summary and Outlook} \label{sec:Summary}
This chapter has compared three established schemes for economic NMPC
\begin{itemize}
\item[(i)] ~~~$V_\mathrm{f}(x) = 0$ \quad \quad~ and \qquad $\mbb{X}_\mathrm{f} = \{\bar x\}$;
\item[(ii)]~~~$V_\mathrm{f}(x) = 0$ \quad \quad~ and \qquad $\mbb{X}_\mathrm{f} = \mbb{R}^{n_x}$;
\item[(iii)]~~~$V_\mathrm{f}(x) = \bar\lambda^\top x$ \quad \,and \qquad $\mbb{X}_\mathrm{f} = \mbb{R}^{n_x}$;
\end{itemize}
 with respect to different aspects: 
\begin{itemize}
\item the role of primal and dual terminal constraints;
\item the characteristics of the feasible set;
\item the length of the stabilizing horizon.
\end{itemize} 
On the considered example, the stabilizing horizon length improves from $N=7$ to $N=1$ from scheme (i) to scheme (iii).
Hence, it turns out that the simple linear end penalty $V_\mathrm{f}(x) = \bar\lambda^\top x$ does not only help with respect to convergence to the optimal steady state, but also with respect to feasibility of the OCP for short horizon. Indeed, the numerical analysis of Section \ref{sec:Examples} indicates that the effect of this terminal constraint on the minimal stabilizing horizon length is substantial as compared to the scheme with terminal constraint. 

Notably, the results of~\cite{kit:faulwasser18e_2} and \cite{kit:faulwasser19c} have also shown that optimal steady-state adjoint $\bar\lambda$ does not need to be known a-priori. In case of accurate models it can be inferred from open-loop predictions \cite{kit:faulwasser18e_2} and in case of model deficiencies it can be estimated from plant data~\cite{kit:faulwasser19c}. However, a formal proof of supremacy in terms of closed-loop performance is currently not available. Moreover, the analysis of Section \ref{sec:Examples} should be validated also on more challenging examples. Finally, we remark that the time-varying counterpart for the scheme with $V_\mathrm{f}(x) = \bar\lambda^\top x$ is still to be explored. Moreover, the preliminary analysis in terms of stabilizing horizon lengths for the different schemes could be extended using 
viability kernel concepts, see e.g. \cite{Boccia14a}.


%
%
\bibliographystyle{plain}
\bibliography{literature_latin1}

\end{document}